\documentclass[preprint,showpacs,preprintnumbers,amsmath,amssymb,nofootinbib,showkeys,aps]{revtex4}
\pdfoutput=1

\usepackage{epsfig}
\usepackage{graphicx}
\usepackage{dcolumn}
\usepackage{bm}
\usepackage{amsmath}
\usepackage{amssymb}
\usepackage{latexsym}
\usepackage{color}
\usepackage{hyperref}
\usepackage{mathrsfs}
\usepackage{float}

\newcommand{\be}{\begin{equation}}
\newcommand{\ee}{\end{equation}}
\newcommand{\bea}{\begin{eqnarray}}
\newcommand{\eea}{\end{eqnarray}}
\newcommand{\texpdf}{\texorpdfstring}

\begin{document}

\title{Cosmological Constraints on $\Lambda(t)$CDM Models}

\author{H. A. P. Macedo$^{1}$} \email{ha.macedo@unesp.br}
\author{L. S. Brito$^{1}$} \email{lucas.s.brito@unesp.br}
\author{J. F. Jesus$^{1,2}$}\email{jf.jesus@unesp.br}
\author{M. E. S. Alves$^{1,3}$}\email{marcio.alves@unesp.br}

\affiliation{$^1$Universidade Estadual Paulista (UNESP), Faculdade de Engenharia e Ciências, Departamento de F\'isica - Av. Dr. Ariberto Pereira da Cunha 333, 12516-410, Guaratinguet\'a, SP, Brazil
\\$^2$Universidade Estadual Paulista (UNESP), Instituto de Ciências e Engenharia - R. Geraldo Alckmin, 519, 18409-010, Itapeva, SP, Brazil
\\$^3$ Universidade Estadual Paulista (UNESP), Instituto de Ciência e Tecnologia, São José dos Campos, SP, 12247-004, Brazil
}

\def\zt{\mbox{$z_t$}}

\begin{abstract}
Problems with the concordance cosmology $\Lambda$CDM as the cosmological constant problem, coincidence problems and Hubble tension has led to many proposed alternatives, as the $\Lambda(t)$CDM, where the now called $\Lambda$ cosmological term is allowed to vary due to an interaction with pressureless matter. Here, we analyze one class of these proposals, namely, $\Lambda=\alpha'a^{-2}+\beta H^2+\lambda_*$, based on dimensional arguments. Using SNe Ia, cosmic chronometers data plus constraints on $H_0$ from SH0ES and Planck satellite, we constrain the free parameters of this class of models. By using the Planck prior over $H_0$, we conclude that the $\lambda_*$ term can not be discarded by this analysis, thereby disfavouring models only with the time-variable terms. The SH0ES prior over $H_0$ has an weak evidence in this direction. The subclasses of models with $\alpha'=0$ and with $\beta=0$ can not be discarded by this analysis. Finally, by using distance priors from CMB, the $\Lambda$ time-dependence was quite restricted.
\end{abstract}

\maketitle

\section{Introduction} 

The concordance cosmological model $\Lambda$CDM ($\Lambda$ plus Cold Dark Matter) is very successful in explaining a variety of cosmological observations as, for instance, the accelerating expansion of the Universe and the power spectrum of the cosmic microwave background radiation (CMB). However, the model suffers several theoretical and observational difficulties. Some remarkable examples are the cosmological constant problem, the coincidence problem, and the Hubble tension (see, e.g., \cite{Perivolaropoulos2022} for a review).  

In the last decades, we have seen an increasing number of alternatives to the $\Lambda$CDM model aiming to alleviate such difficulties. These alternatives range from the quest for dark energy models to extended theories of gravity. In this context, it is natural to investigate if $\Lambda$ is a function of the cosmic time $t$. 

Models with a time-varying $\Lambda$ or Vacuum-decay models have been conceived in different contexts. In several models, some ad hoc time dependence for $\Lambda(t)$ is assumed. Some of the most common examples were addressed in Refs. \cite{Vishwakarma2001, Ahmet2018} (see \cite{Overduin1998} and references therein for a list of phenomenological decay laws of $\Lambda(t)$). The functional form of $\Lambda(t)$ can also be derived, for instance, by geometrical motivations \cite{Azri2012, Azri2017} or from Quantum Mechanical arguments \cite{Szydlowski2015}. The interaction of vacuum with matter has also been considered in different approaches and confronted with recent cosmological data (see, e.g.,  \cite{Benetti2019, Papagian2020, Benetti2021, Bruni2022}). 

A promising approach to overcome the puzzles of the $\Lambda$CDM model is known as the `running vacuum model' (RVM). It emerges when one uses the renormalization group approach of quantum field theory in curved spaces to renormalize the vacuum energy density. It is possible to show that vacuum energy density evolves as a series of powers of the Hubble function $H$ and its derivatives with respect to cosmic time: $\rho_{\rm vac}(H, \dot{H},\ldots)$. The leading term of the expansion is constant, but the next-to-leading one evolves as $H^2$. There are other terms in the expansion that can be relevant for the early Universe cosmology, but the term $H^2$ can affect the current evolution of the scale factor. Initially, the RVM was introduced in a semi-qualitative way through the renormalization group approach. Some of the first motivations of this model can be found, e.g., in the Refs. \cite{Shapiro2000, Babic2001, Shapiro2002, Shapiro2003} (see also \cite{Sola2013} for an old review on the subject). However, in recent years, the RVM was derived from a rigorous analysis within the Quantum Field Theory in curved spacetime. The derivation of the final form of RVM can be found in the Refs. \cite{Pulido2020,Pulido2022a,Pulido2022b,Pulido2023} (see \cite{Peracaula2022} for a review on recent theoretical developments). Moreover, such a class of models can be more favored than the $\Lambda$CDM model when a fit with the cosmological observables is performed \cite{Sola2017, Peracaula2018, Tsiapi2019, Mavromatos2021, Peracaula2023, Peracaula2021, Peracaula2018b}.

To answer the conundrum of why the cosmological constant is so small today, one could also propose models with $\Lambda \propto a^{-m}$, where $a$ is the scale factor and $m$ is a positive constant to be determined. From dimensional arguments by quantum cosmology, it is natural to choose $m = 2$ \cite{Lopez1996, Chen1990}. Therefore, in this perspective, $\Lambda$ has the same decay behavior as the curvature term. Such an evolution of $\Lambda$ was first proposed by Özer and Taha \cite{Ozer1986, Ozer1987} as a way to solve the cosmological problems of the eighties decade.

On the other hand, following similar phenomenological arguments, in Ref. \cite{Carvalho1992}, the authors parametrized the time evolution of $\Lambda$ as the sum of a term proportional to $a^{-2}$ to a term proportional to $H^2$, i.e., the same term that emerges from the RVM. In the present article, we follow this approach, but we add a ``bare'' cosmological term $\lambda_*$. Specifically, we consider four models of a time-varying $\Lambda(t)$ in the class 

\begin{equation}\label{class of models lambda}
    \Lambda_g = \frac{\alpha^\prime}{a^2} + \beta H^2 + \lambda_*,
\end{equation}
with $\alpha^\prime$, $\beta$ and $\lambda_*$ constants. Three models are chosen by selecting one of these constants to vanish identically (these three models are depicted in Table \ref{Tab:3models}) and the fourth model is the complete one for which the three constants are non-null. 

The phenomenological model described by the complete model presents a smooth transition from the early de Sitter stage to the radiation phase. Such a transition is independent of the curvature parameter and solves naturally the horizon and the graceful exit problem \cite{Lima2015}.

To put constraints on the free parameters of the models, we use the SNe Ia sample consisting of 1048 SNe Ia apparent magnitude measurements from the Pantheon sample \cite{pantheon} and a compilation of 32 cosmic chronometers data of the Hubble parameter, $H(z)$ \cite{MorescoEtAl22}. We have also considered the most up-to-date constraints on $H_0$, namely, the ones from SH0ES \cite{Riess2021} and Planck \cite{Planck2020}.

We organized the article as follows. We describe the Friedmann equations with a time-dependent $\Lambda(t)$-term in Section \ref{sec: Friedmann eqs} (neglecting radiation) and Section \ref{FriedEqsRad} (including radiation). We obtain analytical solutions for $H(z)$ in the class of models given by Eq. (\ref{class of models lambda}) in Sections \ref{sec: Class of models} and \ref{sec: Class of modelsRad}. In Section \ref{sec: analysis}, we constrain the parameters of the models using SNe Ia data, cosmic chronometers data and CMB. In the analysis, we consider separately the constraints on $H_0$ from Planck (Section \ref{sec: analysis I}) and from SH0ES (Section \ref{sec: analysis II}). Finally, we present our conclusions and final remarks in Section \ref{sec: conclusion}.

\section{\label{sec: Friedmann eqs} Cosmological equations for a varying \texpdf{\boldmath{$\Lambda$}}{L} term, neglecting radiation}

From the Cosmological Principle and the Einstein Field Equations, we have the so-called Friedmann equations, given by
\begin{align}
H^2 &= \frac{8\pi G \rho_T}{3} - \frac{k}{a^2},\\
\frac{\ddot{a}}{a} &= -\frac{4\pi G}{3}(\rho_T + 3p_T),
\label{fried}
\end{align}
where $\rho_T$ is the total density of the Universe matter-energy content, $p_T$ is total pressure and $k$ is the curvature scalar. As we are mainly interested in the late-time Universe, we shall neglect the radiation contribution, in such a way that $\rho_T$ is given by

\be
\rho_{T} = \rho_{M} + \rho_{\Lambda},
\ee

where $\rho_M$ corresponds to the total pressureless matter (dark matter+baryons) and $\rho_\Lambda$ corresponds to the time-varying $\Lambda(t)$-term. In the present article we assume the equation of state (EoS) of vacuum to be exactly $w_{vac} = -1$ such that $p_\Lambda=-\rho_\Lambda$. However, a recent result for the RVM is that the EoS of vacuum evolves with the cosmic history \cite{Pulido2022b}. This would change our results and may be considered in future works. From the continuity equation, we have
\begin{align}
\dot{\rho}_{M} + 3H{\rho_{M}} &= Q,\\
\dot{\rho}_{\Lambda} &= -Q,
\end{align}
where $Q$ is the interaction term between pressureless matter and vacuum. With these components, the Friedmann equations \eqref{fried} now read
\begin{align}
H^2  &= \frac{8\pi G (\rho_M + \rho_\Lambda)}{3} - \frac{k}{a^2}\label{eqfried},\\
\frac{\ddot{a}}{a} &= -\frac{4\pi G}{3}(\rho_M -2\rho_\Lambda)\label{a2p}.
\end{align}

By multiplying the Eq. \eqref{a2p} by 2, we have
\be
2\frac{\ddot{a}}{a} = -\frac{8\pi G}{3}(\rho_M -2\rho_\Lambda) = -\frac{8\pi G}{3}\rho_M +\frac{16\pi G}{3}\rho_\Lambda\label{2a2p},
\ee
and summing the Eq. \eqref{eqfried} with the Eq. \eqref{2a2p}, we have
\bea
H^2  + \frac{2\ddot{a}}{a}&=& 8\pi G\rho_\Lambda - \frac{k}{a^2}.
\label{eqH2}
\eea
Since $\Lambda = 8\pi G\rho_\Lambda$, Eq. \eqref{eqH2} reads
\be
H^2 = -\frac{2\ddot{a}}{a} + \Lambda - \frac{k}{a^2}.
\ee
By replacing $\frac{\ddot{a}}{a}=\dot{H}+H^2$, we find
\bea
3H^2  &=& -2\dot{H} + \Lambda - \frac{k}{a^2}.
\eea

In order to perform cosmological constraints, let us now change to derivatives with respect to the redshift
\bea
\frac{d}{dt} = -H(1+z)\frac{d}{dz}.
\eea

Thus, the equation
\bea
2\dot{H}  &=& -3H^2 + \Lambda - \frac{k}{a^2},
\eea
now reads
\bea
-2H(1+z)\frac{dH}{dz} &=& -3H^2 + \Lambda - \frac{k}{a^2}, 
\eea
and by replacing $k = -\Omega_{k0}H_0^2$ we have
\bea
\frac{dH}{dz} = \frac{3H}{2(1+z)} - \frac{\Omega_{k0} H_0^2(1+z)}{2H} - \frac{\Lambda}{2H (1+z)}.
\eea

If we further use the definition $E \equiv \frac{H}{H_0}$ the above equation reads
\bea
\frac{dE}{dz} &=& \frac{3E}{2(1+z)} - \frac{\Omega_{k0}(1+z)}{2E} - \frac{\Lambda}{2E{H_0^2}(1+z)}\label{dEdz}.
\eea

From now on, we shall assume that the Universe is spatially flat ($k = 0$), as indicated by inflation and CMB. Therefore, we finally obtain the equation
\bea
\frac{dE}{dz} &=& \frac{3E}{2(1+z)} - \frac{\Lambda}{2E{H_0^2}(1+z)}.
\label{dEdzk0}
\eea

For a given $\Lambda(a,H)$ (or $\Lambda(z,H)$), Eq. \eqref{dEdzk0} can be solved in order to obtain the universe evolution $E(z)$. In the next subsection, we shall assume a fair general $\Lambda(a,H)$ dependence in order to solve this equation and compare the assumed models with cosmological observations.

\subsection{\texpdf{\boldmath{$\Lambda = \alpha' a^{-2} + \beta {H}^2 + \lambda_*$}}{L=alpha/(a*a)+beta*H*H} class of models, neglecting radiation}\label{sec: Class of models}

Dimensional arguments have led to the proposals of $\Lambda\propto a^{-2}$, $\Lambda\propto H^2$ models in the literature. Here we test a combination of these proposals together with a constant term, in order to find which of these terms may contribute the most to the evolution of the universe as indicated by observations. So, the models we study here are derived from the following $\Lambda$ dependence
\be
\Lambda_{g} = \frac{\alpha'}{a^2} + \beta {H}^2 + \lambda_*\label{lgeral}.\\
\ee

We shall not consider this general $\Lambda_g$ as a model to be constrained by observations, as it has too many free parameters, and it may be penalized in a Bayesian criterion. Actually, we choose to work with particular cases of this $\Lambda_g$ dependence, where, in each case, one parameter contribution is neglected, as summarised in Table \ref{Tab:3models}

\begin{table}[!ht]
    \centering
    \begin{tabular}{c|c | c}
    
              ~~~Model~~~  &~~~$\Lambda$~~~&~~~ Fixed parameter~~~ \\
                \hline
     $\Lambda_1$    &  $\frac{\alpha'}{a^2} + \beta {H}^2$ &  $\lambda_*=0$        \\
     $\Lambda_2$      &  $\frac{\alpha'}{a^2} + \lambda_*$ &     $\beta=0$         \\
     $\Lambda_3$  &     $\beta {H}^2 + \lambda_*$ & $\alpha'=0$   \\
    
    \end{tabular}
    \caption{Here we summarise the three models considered in the article.}
    \label{Tab:3models}
\end{table}

Let us now obtain the evolution of this class of models. From Eq. \eqref{lgeral}, we have the values today
\be
\Lambda_0=\alpha'+\beta H_0^2+\lambda_*.
\ee

As $\Omega_{\Lambda}\equiv\frac{\Lambda_0}{3H_0^2}$, we may also write
\be
\Omega_{\Lambda} = \frac{\alpha}{3}+\frac{\beta}{3}+\frac{\lambda_*}{3H_0^2} = \frac{\alpha}{3}+\frac{\beta}{3}+\Omega_{\lambda*},
\ee
where we have defined $\Omega_{\lambda*}\equiv\frac{\lambda_*}{3H_0^2}$ and the dimensionless $\alpha\equiv\frac{\alpha'}{H_0^2}$, for mathematical convenience. From this, we may write for $\alpha$
\be
\alpha =  3\left(\Omega_{\Lambda}-\frac{\beta}{3}-\Omega_{\lambda*}\right).
\ee

As already mentioned, we choose to work with a spatially flat Universe, such that from Eq. \eqref{eqfried}, we have the normalization condition
\be
\Omega_{\Lambda}=1-\Omega_m.
\ee

Now, with these dimensionless parameters $(\Omega_m,\beta,\Omega_{\lambda*})$, an analytical solution for the Eq. \eqref{dEdzk0} can be obtained. The general solution is given by  
\begin{align}
    E^2(z) &= \left[\frac{6 \Omega_{\lambda_*} + (3-\beta ) (1-3\Omega_{\Lambda})}{(1-\beta ) (3-\beta )}\right](1 + z)^{3 - \beta} + \frac{3\Omega_{\lambda_*}}{3 - \beta}+\left[\frac{- 3\Omega_{\lambda_*} + 3\Omega_{\Lambda} -\beta}{1 - \beta}\right](1 + z)^2.
\end{align}

The solutions for each one of the three models depicted in the Table \ref{Tab:3models} are particular cases of this solution obtined by the appropriate choice of parameters. They are given by
\begin{itemize}
    \item Model $\Lambda_1 =  \frac{\alpha'}{a^2} + \beta {H}^2$, ($\lambda_*=0$) 
    \bea
    E(z) &=& \sqrt{\frac{1 - 3\Omega_{\Lambda}}{1 - \beta}(1 + z)^{(3 - \beta)} + \frac{3\Omega_{\Lambda}-\beta}{1 - \beta} (1 + z)^2};
    \eea
    \item Model $\Lambda_2 =  \frac{\alpha'}{a^2} + \lambda_*$, ($\beta=0$)
    \bea
    E(z) &=& \sqrt{(1 + 2\Omega_{\lambda_*} - 3\Omega_{\Lambda})(1 + z)^3 + \Omega_{\lambda_*} + 3(-\Omega_{\lambda_*} + \Omega_{\Lambda})(1 + z)^2 };
    \label{Ez for the lambda2 model}
    \eea
    \item Model $\Lambda_3 =  \beta {H}^2 + \lambda_*$, ($\alpha^\prime=0$)
    \be
    E(z) = \sqrt{\left[\frac{6 \Omega_{\lambda_*} + (3-\beta ) (1-3\Omega_{\Lambda})}{(1-\beta ) (3-\beta )}\right](1 + z)^{3 - \beta} + \frac{3\Omega_{\lambda_*}}{3 - \beta}}.
    \ee
\end{itemize}

It is worth noticing that, for the $\Lambda_2$ model, the $E(z)$ given by Eq. (\ref{Ez for the lambda2 model}) is similar to the $\Lambda$CDM model with spatial curvature. This is due to the fact that the term $\propto a^{-2}$ mimics a curvature term in this case. 

The functions $E(z)$ we have obtained are all we need in order to constrain the three models with observational data in the next section. We can also obtain the interaction term for each model, in order to analyze its behavior later. For the general case \eqref{lgeral}, we have the following interaction term

\be
\mathcal{Q}(z)\equiv\frac{8 \pi G}{H_0^3}Q(z) = 2\alpha(1+z)^2E(z) + \beta E(z)(1+z) \frac{dE^2(z)}{dz}.
\label{QzGeral}
\ee

\section{Cosmological Equations for a Varying \texorpdfstring{\boldmath{$\Lambda$}}{Lambda} Term, including  radiation}\label{FriedEqsRad}
Taking radiation into account, the Friedmann equations \eqref{fried} are the same, but now we have $\rho_T$ and $p_T$ given by:
\begin{align}
\rho_{T} &= \rho_{M} + \rho_r+\rho_{\Lambda}\\
p_T &= p_r+p_\Lambda=\frac{\rho_r}{3}-\rho_\Lambda
\end{align}
where $\rho_r$ is radiation density and $\rho_M=\rho_d+\rho_b$  (dark matter ($d$)+baryons ($b$)). The continuity equations now read
\begin{align}
\dot{\rho}_{b} + 3H{\rho_{b}} &= 0,\label{rhob}\\
\dot{\rho}_{r} + 4H{\rho_{r}} &= 0,\label{rhor}\\
\dot{\rho}_d + 3H{\rho_d} &= Q,\label{rhod}\\
\dot{\rho}_{\Lambda} &= -Q,
\end{align}
where $Q$ is the interaction term between dark matter and vacuum. It is interesting to note that Eqs. \eqref{rhob} and \eqref{rhod} can be combined to write a continuity equation for total pressureless matter:
\be
\dot{\rho}_M + 3H{\rho_M} = Q
\ee
With these components, the Friedmann equations \eqref{fried} now read
\begin{align}
H^2  &= \frac{8\pi G (\rho_M + \rho_r + \rho_\Lambda)}{3} - \frac{k}{a^2}\label{eqfriedr},\\
\frac{\ddot{a}}{a} &= -\frac{4\pi G}{3}(\rho_M +2\rho_r - 2\rho_\Lambda)\label{a2pr}.
\end{align}

It can be shown, that following the same steps as in Sec. II, we may arrive at the general result, with spatial curvature:
\bea
\frac{dE}{dz} &=& \frac{3E}{2(1+z)} - \frac{\Omega_{k0}(1+z)}{2E} +\frac{\Omega_{r0}(1+z)^3}{2E}- \frac{\Lambda}{2E{H_0^2}(1+z)}\label{dEdzRad}.
\eea

And, by assuming that the Universe is spatially flat ($k = 0$), as indicated by inflation and CMB, we obtain the equation:
\bea
\frac{dE}{dz} &=& \frac{3E}{2(1+z)} +\frac{\Omega_{r0}(1+z)^3}{2E} - \frac{\Lambda}{2E{H_0^2}(1+z)}.
\label{dEdzk0Rad}
\eea

For a given $\Lambda(a,H)$ (or $\Lambda(z,H)$), Eq. \eqref{dEdzk0Rad} can be solved in order to obtain the universe evolution $E(z)$. In the next subsection, we  assume the same $\Lambda(a,H)$ dependence as before \eqref{lgeral} in order to solve this equation and compare the assumed models with cosmological observations.

\subsection{\texpdf{\boldmath{$\Lambda = \alpha' a^{-2} + \beta {H}^2 + \lambda_*$}}{L=alpha/(a*a)+beta*H*H} class of models, including radiation}\label{sec: Class of modelsRad}

For this class, including radiation, the normalization condition now reads:
\be
\Omega_{\Lambda}=1-\Omega_m-\Omega_r.
\ee

Now, with the dimensionless parameters $(\Omega_m,\Omega_r,\alpha,\beta,\Omega_{\lambda*})$, an analytical solution for the Eq. \eqref{dEdzk0} can be obtained. The general solution is given by 
\begin{align}
    E^2(z) &= \frac{(1+z)^{3-\beta}}{3-\beta}\left(\frac{2 \alpha }{\beta -1}+3 \Omega_{m0}+\frac{4 \beta  \Omega_{r0}}{1+\beta}\right)+\frac{\Omega_{r0}(1+z)^4}{(1+\beta)}+\frac{\alpha(1+z)^2}{(1-\beta)}+\frac{3 \Omega_{\lambda*}}{3-\beta }
\end{align}
which is a general solution in the cases that $\beta\notin\{-1,1,3\}$.

\section{\label{sec: analysis} Analysis and Results}
For this analysis, we use 3 variations of the general equation, being first with $\lambda_* = 0$, second with $\beta = 0$, and last we take $\alpha = 0$, as described in Tab. \ref{Tab:3models}.

In order to constrain the models in the present work, we have used as observational data, the SNe Ia sample consisting of $1048$ SNe Ia apparent magnitude measurements from the Pantheon sample \cite{pantheon} and a compilation of $32$ Hubble parameter data, $H(z)$ \cite{MorescoEtAl22}, obtained by estimating the differential ages of galaxies, called Cosmic Chronometers (CCs).

The 32 $H(z)$ CCs data is a sample compiled by \cite{MorescoEtAl22}, consisting of $H(z)$ data within the range $0.07 < z < 1.965$. In the Ref. \cite{MorescoEtAl22}, the authors have estimated systematic errors for these data, by running simulations and considering effects such as metallicity, rejuvenation effect, star formation history, initial mass function, choice of stellar library etc.\footnote{The method to obtain the full covariance matrix, together with jupyter notebooks as examples are furnished by M. Moresco at \url{https://gitlab.com/mmoresco/CCcovariance}.}

The Pantheon compilation consists of $1048$ data from SNe Ia, within the redshift range $0.01\,<\, z\,<\,2.3$, containing measurements of SDSS, Pan-STARRS1 (PS1), SDSS, SNLS, and various HST and low-$z$ datasets.

In order to better constrain the models, besides SNe Ia+$H(z)$ data, we have also considered the most up-to-date constraints over $H_0$, namely, the ones from SH0ES $(73.2 \pm 1.3)$ km/s/Mpc \cite{Riess2021} and Planck $(67.36 \pm 0.54)$ km/s/Mpc \cite{Planck2020}.  As it is well known, these constraints are currently in conflict, generating the so-called ``$H_0$ tension'' \cite{H0TensionReview}. It is important to mention that these constraints are obtained from quite different methods. While the SH0ES $H_0$ is obtained simply from local constraints, following the distance ladder built from Cepheid distances and local SNe Ia, the Planck $H_0$ is obtained from high redshift constraints, assuming the flat $\Lambda$CDM model. Given this tension, we preferred to make two separate analyses, one considering the $H_0$ from Planck and one taking into account the $H_0$ from SH0ES.

Below, we show the results of our analyses, first showing the constraints from SNe Ia+$H(z)$+$H_0$ from Planck, and later showing the constraints from SNe Ia+$H(z)$+$H_0$ from SH0ES.

In all the analyses that we have made, we have assumed the flat priors over the free parameters: $\alpha\in[-2,2]$, $\beta\in[-5,15]$, $\Omega_m\in[0,1]$, $H_0\in[55,85]$ km/s/Mpc. It is important to note that while SNe Ia data constrain $\{\Omega_m$, $\alpha$, $\beta\}$, $H(z)$ data constrain $\{H_0, \Omega_m, \alpha, \beta\}$. However, $H(z)$ data alone do not provide strong constraints over the free parameters, so we choose to work with $H(z)$+SNe Ia data combination. Furthermore, we have added constraints over $H_0$ from Planck and SH0ES in order to probe the $H_0$ tension and also because they consist of strong constraints over $H_0$.

\subsection{SNe Ia+\texpdf{\boldmath{$H(z)$}}{Hz}+Planck \texpdf{\boldmath{$H_0$}}{H0} analysis}\label{sec: analysis I}

\begin{figure}[H] 
\begin{center}
    \includegraphics[width=0.49\textwidth]{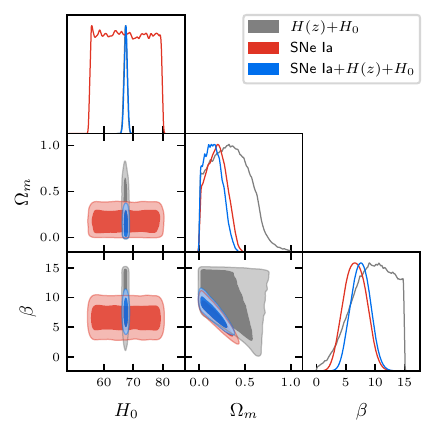}
	\includegraphics[width=0.49\textwidth]{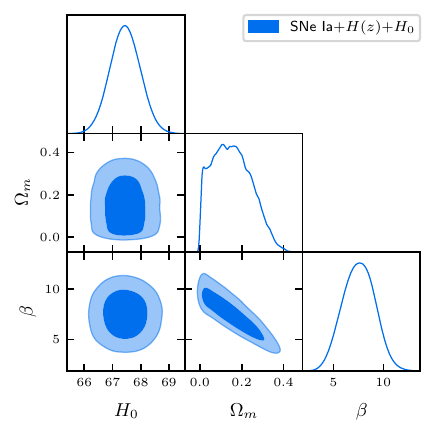}
	\caption{SNe Ia, $H(z)$ and Planck $H_0$ constraints for $\lambda_* = 0$ ($\Lambda_1$ model), at 1 and 2$\sigma$ c.l., $H_0$ units are km/s/Mpc. \textbf{Left:} SNe Ia, $H(z)$+Planck $H_0$ and SNe Ia+$H(z)$+Planck $H_0$ constraints. \textbf{Right:} Joint constraints from SNe Ia+$H(z)$+Planck $H_0$.}
 \label{lambda*0Pl}
\end{center}
\end{figure}

We start with the $\Lambda_1$ model (see Table \ref{Tab:3models}). As one may see in Fig. \ref{lambda*0Pl} (left), $H_0$ is well constrained by $H(z)$+Planck $H_0$ data, but $\Omega_m$ and $\beta$ are poorly constrained. In fact, one may see that $\Omega_m$ is constrained by the prior in its inferior limit, while $\beta$ is constrained by the prior in its superior limit. $H_0$ can not be constrained by SNe Ia, but $\beta$ is well constrained by this data. $\Omega_m$ is better constrained by SNe Ia, but also is constrained inferiorly by the prior. In Fig. \ref{lambda*0Pl} (right), we can see the result for the joint analysis, where $\Omega_m$ and $\beta$ are better constrained, although $\Omega_m$ yet is constrained inferiorly by the prior. We show the best-fit values for the parameters of the $\Lambda_1$ model in Table \ref{tabPlL1}.

\begin{table}[H]
    \centering
    \begin{tabular} { c  c}

 Parameter &  95\% limits\\
\hline
{\boldmath$H_0            $ \textbf{(km/s/Mpc)}} & $67.4\pm1.1        $\\

{\boldmath$\Omega_m       $} & $0.15^{+0.16}_{-0.15}      $\\

{\boldmath$\beta          $} & $7.5^{+3.2}_{-3.1}         $\\
\hline
\end{tabular}
    \caption{SNe Ia+$H(z)$+Planck $H_0$ constraints for $\lambda_* = 0$ ($\Lambda_1$ model)}
    \label{tabPlL1}
\end{table}

\begin{figure}[H] 
\begin{center}
    \includegraphics[width=0.49\textwidth]{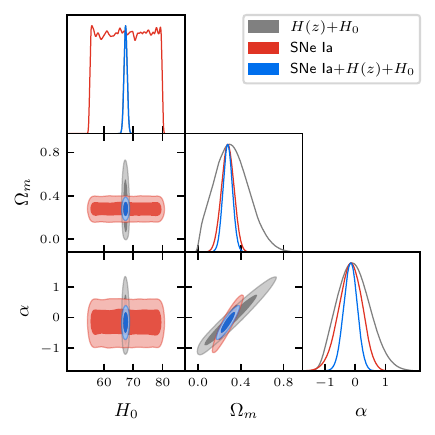}
    \includegraphics[width=0.49\textwidth]{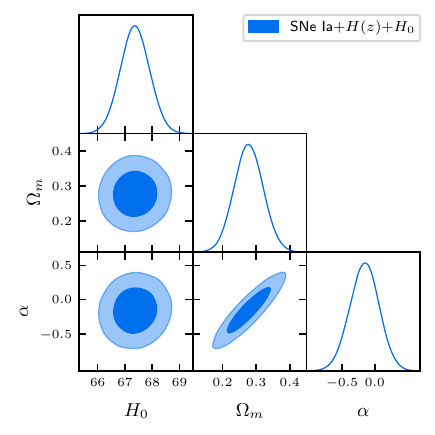}
	\caption{SNe Ia, $H(z)$ and Planck $H_0$ constraints for $\beta = 0$ ($\Lambda_2$ model), at 1 and 2$\sigma$ c.l., $H_0$ units are km/s/Mpc. \textbf{Left:} SNe Ia, $H(z)$+Planck $H_0$ and SNe Ia+$H(z)$+Planck $H_0$ constraints. \textbf{Right:} Joint constraints from SNe Ia+$H(z)$+Planck $H_0$.}
 \label{beta0Pl}
\end{center}
\end{figure}

In Fig. \ref{beta0Pl} we show the analysis for the $\Lambda_2$ model ($\beta = 0$). As one can see in the left panel, $H(z)$+Planck $H_0$ data constrains well $H_0$ and $\alpha$, but not $\Omega_m$, which is constrained by the prior in its inferior limit. As SNe Ia does not constrain $H_0$, but $\Omega_m$ and $\alpha$ are well constrained by this data, it is interesting to combine SNe Ia and $H(z)$ data. One may see in Fig. \ref{beta0Pl} (Right) and Table \ref{tab: constraints for the lambda2 model}, the result for the joint analysis, where $H_0$, $\Omega_m$ and $\alpha$ are better constrained.

\begin{table}[H]
    \centering
    \begin{tabular} { c  c}

 Parameter &  95\% limits\\
\hline
{\boldmath$H_0            $ \textbf{(km/s/Mpc)}} & $67.4\pm1.1        $\\

{\boldmath$\Omega_m       $} & $0.278^{+0.088}_{-0.086}   $\\

{\boldmath$\alpha         $} & $-0.16^{+0.44}_{-0.45}     $\\
\hline
\end{tabular}
    \caption{SNe Ia+$H(z)$+Planck $H_0$ constraints for $\beta = 0$ ($\Lambda_2$ model)}
    \label{tab: constraints for the lambda2 model}
\end{table}

\begin{figure}[H] 
\begin{center}
    \includegraphics[width=0.49\textwidth]{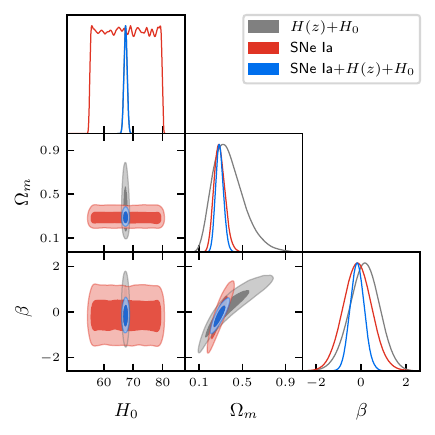}
	\includegraphics[width=0.49\textwidth]{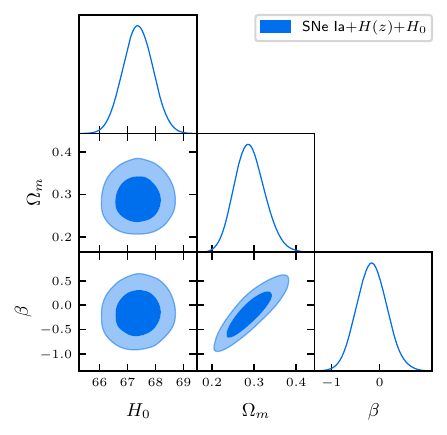}
	\caption{SNe Ia, $H(z)$ and Planck $H_0$ constraints for $\alpha = 0$ ($\Lambda_3$ model), at 1 and 2$\sigma$ c.l., $H_0$ units are km/s/Mpc. \textbf{Left:} SNe Ia, $H(z)$+Planck $H_0$ and SNe Ia+$H(z)$+ Planck $H_0$ constraints. \textbf{Right:} SNe Ia+$H(z)$+Planck $H_0$ joint constraints.}
 \label{alpha0Pl}
\end{center}
\end{figure}

Next, we analyze the $\Lambda_3$ model ($\alpha = 0$) for which we also put constraints on the parameters $\{H_0, \Omega_m, \beta\}$. In Fig. \ref{alpha0Pl} (Left), we can see that $H_0$ and $\beta$ are well constrained by $H(z)$+Planck $H_0$ data, while $\Omega_m$ is weakly constrained.  $\Omega_m$ and $\beta$ are well constrained by SNe Ia, thus complementing the $H(z)$+Planck $H_0$ data constraints. In Fig. \ref{alpha0Pl} (Right) and Table \ref{tab: constraints for the lambda3 model}, we highlight the result for the joint analysis, where $H_0$, $\Omega_m$, and $\beta$ are better constrained.

\begin{table}[H]
    \centering
    \begin{tabular} { c  c}

 Parameter &  95\% limits\\
\hline
{\boldmath$H_0            $ \textbf{(km/s/Mpc)}} & $67.4\pm1.1        $\\

{\boldmath$\Omega_m       $} & $0.290^{+0.076}_{-0.067}   $\\

{\boldmath$\beta          $} & $-0.16^{+0.64}_{-0.61}     $\\
\hline
\end{tabular}
    \caption{SNe Ia+$H(z)$+Planck $H_0$ constraints for $\alpha = 0$ ($\Lambda_3$ model)}
    \label{tab: constraints for the lambda3 model}
\end{table}

\begin{figure}[H] 
\begin{center}
   \includegraphics[width=0.49\textwidth]{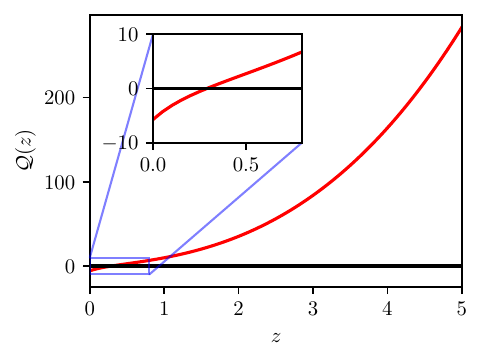}\\
   \includegraphics[width=0.49\textwidth]{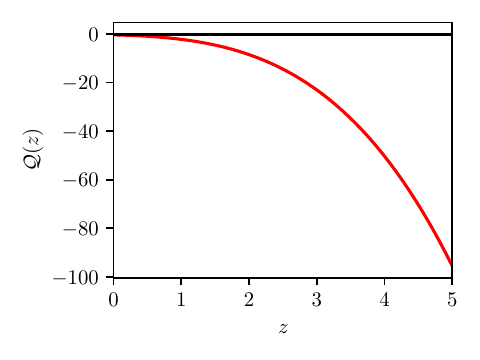}
   \includegraphics[width=0.49\textwidth]{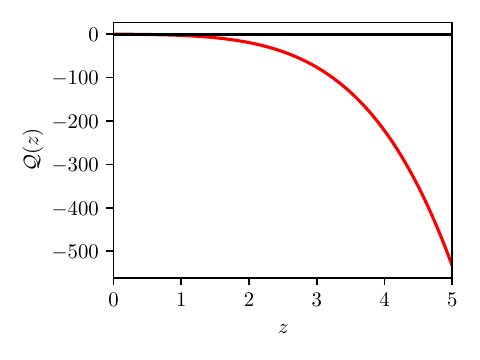}
   \caption{Interaction Term $\mathcal{Q}(z)$ for the best fit parameters from SNe Ia+$H(z)$+$H_0$ from Planck. \textbf{Upper panel:} $\lambda_* = 0  $ ($\Lambda_1$ model). \textbf{Bottom Left:} $\beta = 0  $ ($\Lambda_2$ model). \textbf{Bottom right:} $\alpha = 0  $ ($\Lambda_3$ model).}
 \label{InteractionPL}
\end{center}
\end{figure}

From Fig. \ref{InteractionPL} (upper panel), one may notice that in the past ($z\gtrsim0.3$) the interaction term was positive, meaning a vacuum decaying into DM. However, it is interesting that, for this model ($\Lambda_1$), the interaction term changes sign at low redshift, indicating that now we have decaying of DM into $\Lambda$. This is due to the fact that $\beta>0$ and $\alpha<0$ in the best fit, leading to a change of sign of $\mathcal{Q}(z)$, as one may see from Eq. \eqref{QzGeral}. For $\Lambda_2$ and $\Lambda_3$ models, however, the interaction term is always negative, thus indicating a decaying of DM into $\Lambda$. We may conclude that, at least for the best-fit models, that $\Lambda_1$ is the only model that alleviates or solves the Cosmological Constant Problem (CCP). However, as one can see from Tables II and III, $\alpha$ and $\beta$ may have positive values within 95\% c.l., thus allowing also for decaying of $\Lambda$ into DM. For $\Lambda_1$, however, there is not such a change of tendency when we change the values of $\alpha$ and $\beta$ within 95\% c.l.

\subsection{SNe Ia+\texpdf{\boldmath{$H(z)$}}{Hz}+SH0ES \texpdf{\boldmath{$H_0$}}{H0} analysis} \label{sec: analysis II}

\begin{figure}[H] 
\begin{center}
    \includegraphics[width=0.49\textwidth]{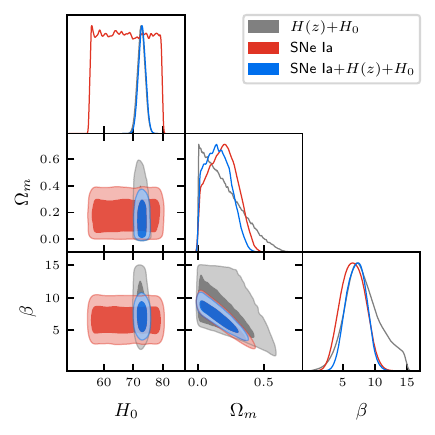}
	\includegraphics[width=0.49\textwidth]{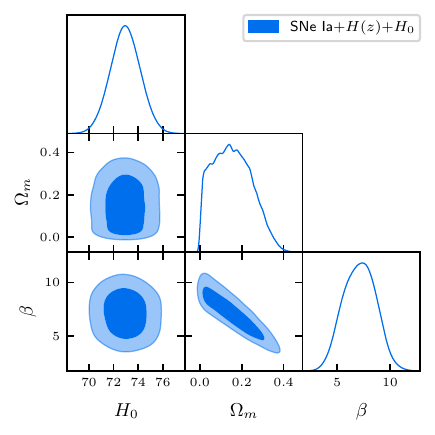}
	\caption{SNe Ia+$H(z)$+$H_0$ from SH0ES for $\lambda_{*} = 0$ ($\Lambda_1$ model), at 1 and 2$\sigma$ c.l., $H_0$ units are km/s/Mpc. \textbf{Left:} SNe Ia, $H(z)$+SH0ES $H_0$ and SNe Ia+$H(z)$+$H_0$ from SH0ES constraints. \textbf{Right:} SNe Ia+$H(z)$+$H_0$ from SH0ES joint constraints.}
 \label{lambda*0SH}
\end{center}
\end{figure}

In Fig. \ref{lambda*0SH}, we may see the constraints from SNe Ia, $H(z)$ data and $H_0$ from SH0ES over the model $\Lambda_1$ (with $\lambda_{*} = 0$).

As one may see in Fig. \ref{lambda*0SH} (left), $H(z)$+SH0ES $H_0$ data constrains well $H_0$, but not $\Omega_m$, as it is being constrained by the prior in its inferior limit, while $\beta$ is constrained by the prior in its superior limit. $H_0$ is not constrained by SNe Ia, but $\Omega_m$ and $\beta$ are better constrained by this data. In Fig. \ref{lambda*0SH} (right), we can see the result for the joint analysis, where $\Omega_m$ and $\beta$ are better constrained, although $\Omega_m$ still is constrained by the prior in its inferior limit. The best-fit values for the $\Lambda_1$ model in this case are shown in Table \ref{tab: constraints to lambda1 model shoes}

\begin{table}[ht]
    \centering
    \begin{tabular} { c  c}

 Parameter &  95\% limits\\
\hline
{\boldmath$H_0            $ \textbf{(km/s/Mpc)}} & $73.0^{+2.4}_{-2.4}        $\\

{\boldmath$\Omega_m       $} & $0.16^{+0.16}_{-0.15}      $\\

{\boldmath$\beta          $} & $7.1^{+3.0}_{-2.9}         $\\
\hline
\end{tabular}
    \caption{SNe Ia+$H(z)$+SH0ES $H_0$ constraints for $\lambda_* = 0$ ($\Lambda_1$ model).}
    \label{tab: constraints to lambda1 model shoes}
\end{table}

\begin{figure}[H] 
\begin{center}
    \includegraphics[width=0.49\textwidth]{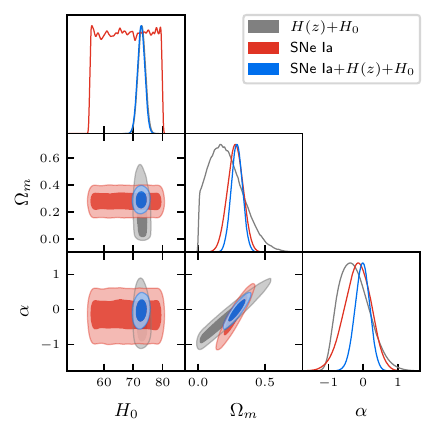}
    \includegraphics[width=0.49\textwidth]{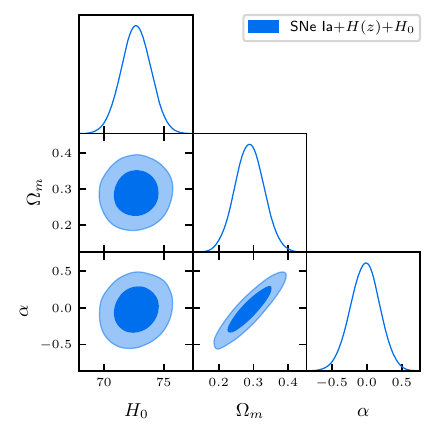}
	\caption{SNe Ia+$H(z)$+$H_0$ from SH0ES for $\beta = 0$ ($\Lambda_2$ model), at 1 and 2$\sigma$ c.l., $H_0$ units are km/s/Mpc. \textbf{Left:} SNe Ia, $H(z)$+SH0ES $H_0$ and SNe Ia+$H(z)$+SH0ES $H_0$ constraints. \textbf{Right:} SNe Ia+$H(z)$ joint constraints.}
 \label{beta0SH}
\end{center}
\end{figure}

In Fig. \ref{beta0SH}, we may see the constraints from SNe Ia, $H(z)$ data and $H_0$ from SH0ES over the model $\Lambda_2$ (with $\beta = 0$).

In Fig. \ref{beta0SH} (left), $H(z)$ data constrains well $H_0$ and $\alpha$, but not $\Omega_m$, which is constrained by the prior in its inferior limit. $H_0$ is not constrained by SNe Ia, but $\Omega_m$ and $\alpha$ are well constrained by this data. In Fig. \ref{beta0SH} (right) and in Table \ref{tab: constraints lambda2 shoes}, we can see the result for the joint analysis, where $\Omega_m$ and $\beta$ are better constrained. We may see that $H_0$, $\Omega_m$ and $\alpha$ are well constrained by this analysis.

\begin{table}[H]
    \centering
    \begin{tabular} { c  c}

 Parameter &  95\% limits\\
\hline
{\boldmath$H_0            $ \textbf{(km/s/Mpc)}} & $72.7^{+2.5}_{-2.5}        $\\

{\boldmath$\Omega_m       $} & $0.289^{+0.085}_{-0.084}   $\\

{\boldmath$\alpha         $} & $-0.03^{+0.42}_{-0.43}     $\\
\hline
\end{tabular}
    \caption{SNe Ia+$H(z)$+ SH0ES $H_0$ constraints for $\beta = 0$ ($\Lambda_2$ model).}
    \label{tab: constraints lambda2 shoes}
\end{table}

Below, in Fig. \ref{alpha0SH}, we may see the constraints from SNe Ia, $H(z)$ data and $H_0$ from SH0ES over the model $\Lambda_3$ (with $\alpha = 0$).

\begin{figure}[H] 
\begin{center}
    \includegraphics[width=0.49\textwidth]{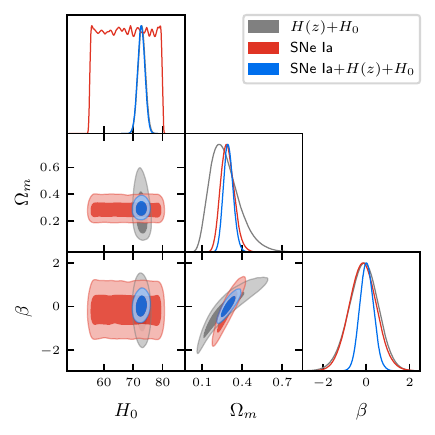}
	\includegraphics[width=0.49\textwidth]{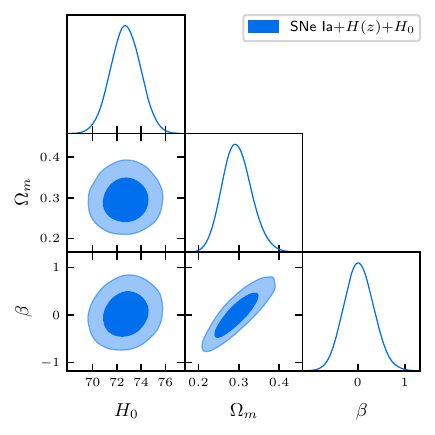}
	\caption{SNe Ia+$H(z)$+$H_0$ from SH0ES for $\alpha = 0$ ($\Lambda_3$ model), at 1 and 2$\sigma$ c.l., $H_0$ units are km/s/Mpc. \textbf{Left:} SNe Ia, $H(z)$+SH0ES $H_0$ and SNe Ia+$H(z)$+SH0ES $H_0$ constraints. \textbf{Right:} SNe Ia+$H(z)$ joint constraints.}
 \label{alpha0SH}
\end{center}
\end{figure}

We can see in Fig. \ref{alpha0SH} (left), $H(z)$+SH0ES $H_0$ data constraints well $H_0$ and $\beta$, but not $\Omega_m$, which is being constrained by the prior in its inferior limit. $H_0$ is not constrained by SNe Ia, but $\Omega_m$ and $\beta$ are well constrained by this data. One may see in Fig. \ref{alpha0SH} (right), the result for the joint analysis, where $H_0$, $\Omega_m$, and $\beta$ are well constrained. In Table \ref{tabSHL3} we show the best-fit values for this model. 

\begin{table}[H]
    \centering
    \begin{tabular} { c  c}

Parameter &  95\% limits\\
\hline
{\boldmath$H_0            $ \textbf{(km/s/Mpc)}} & $72.7^{+2.5}_{-2.5}        $\\

{\boldmath$\Omega_m       $} & $0.296^{+0.078}_{-0.069}   $\\

{\boldmath$\beta          $} & $0.03^{+0.65}_{-0.61}      $\\
\hline
\end{tabular}
    \caption{SNe Ia+$H(z)$+SH0ES $H_0$ constraints for $\alpha = 0$ ($\Lambda_3$ model).}
    \label{tabSHL3}
\end{table}

\begin{figure}[H] 
\begin{center}
   \includegraphics[width=0.49\textwidth]{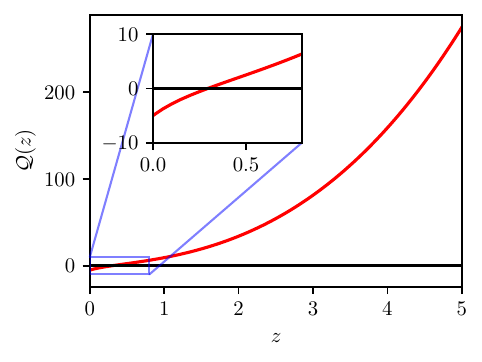}\\
   \includegraphics[width=0.49\textwidth]{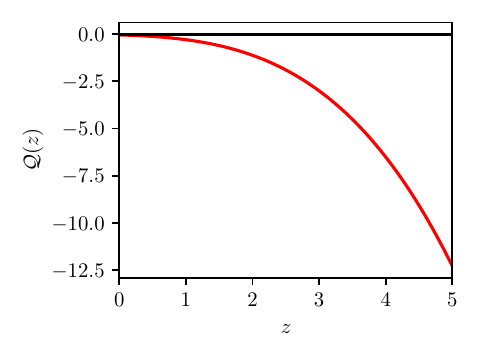}
   \includegraphics[width=0.49\textwidth]{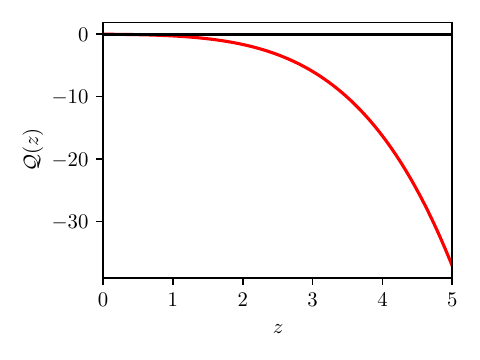}
   \caption{Interaction Term $\mathcal{Q}(z)$ for the best fit parameters from SNe Ia+$H(z)$+SH0ES $H_0$. \textbf{Upper panel:} $\lambda_* = 0  $ ($\Lambda_1$ model). \textbf{Bottom Left:} $\beta = 0  $ ($\Lambda_2$ model). \textbf{Bottom right:} $\alpha = 0  $ ($\Lambda_3$ model).}
 \label{InteractionSH}
\end{center}
\end{figure}

As one may see from Fig. \ref{InteractionSH}, the interaction term has quite similar behavior to Fig. \ref{InteractionPL}. That is, $\mathcal{Q}(z)$ changes sign for the model $\Lambda_1$ and it is negative for models $\Lambda_2$ and $\Lambda_3$, increasing its absolute value with redshift. However, the best-fit parameters from the analysis with the SH0ES $H_0$ prior indicate less interaction in the past than in the case with the Planck $H_0$ prior.

We can see from the figures above, that the most stringent constraints from the data are in the context of the model $\Lambda_3$, followed by the constraints over the model $\Lambda_2$. While the worst constraints over the parameters are in the case of the model $\Lambda_1$. One reason for that is that, as one may see, the contours from SNe Ia and $H(z)+H_0$ are misaligned in the cases $\Lambda_3$ and $\Lambda_2$ while being aligned in the case of $\Lambda_1$. Then, we may say that in the context of the models, $\Lambda_3$ and $\Lambda_2$, the SNe Ia and $H(z)+H_0$ observations nicely complement each other.

Finally, in order to make a more quantitative comparison among the models analyzed here, we use the Bayesian Information Criterion (BIC) to conclude which model better describes the analyzed data. It is important to mention that the BIC takes into account not only the goodness of fit but also penalizes the excess of free parameters, in agreement with the notion of the \textit{Ockham's razor}. Therefore, BIC favors simpler models. BIC can be written as \cite{Schwarz78,Liddle04}
\be
\text{BIC} = -2 \ln\mathcal{L}_{max} + p\ln N,
\ee
where $N$ is the number of data, $\mathcal{L}_{max}$ is the maximum of likelihood and $p$ is the number of free parameters. In the model comparison, the model with achieves the lowest BIC is favored. The likelihood has some normalization, such that instead of using the absolute BIC value, what really is important to consider is the relative BIC among models, which is given by
\be
\Delta\text{BIC} = -(\text{BIC}_i-\text{BIC}_j).
\ee

The level of support for each model depends on the value of the $\Delta$BIC and is explained at \cite{bayesccdm}. As BIC is a model comparison, in Table \ref{tabBIC} below, we calculated $\Delta$BIC relative to $\Lambda_3$, which has the lowest BIC in the case of the Planck $H_0$ prior and relative to $\Lambda_2$, which has the lowest BIC in the case of the SH0ES $H_0$ prior. The level of support can be seen in the last column of this table.

\begin{table}[H]
\centering
\begin{tabular}{|c|c|c|c|c|c|c|c|}
\hline
Model & Data & $\chi_{min}^2$ & $n_{par}$ & $n_{data}$ & BIC &  $\Delta$BIC & Support \\ \hline
$\Lambda_g$ & SNe Ia+$H(z)$+Planck $H_0$ & 1043.085 & 4 & 1080 & 1071.027 & 6.984 & Decisive\\ \hline
$\Lambda_1$ ($\lambda_{*} = 0$) & SNe Ia+$H(z)$+Planck $H_0$ & 1047.614 & 3 & 1080 & 1068.571 & 4.528 & Strong to very strong\\ \hline
$\Lambda_2$ ($\beta = 0$) & SNe Ia+$H(z)$+Planck $H_0$ & 1043.091 & 3 & 1080 & 1064.048 & 0.005 & Not significant\\ \hline
$\Lambda_3$ ($\alpha = 0$) & SNe Ia+$H(z)$+Planck $H_0$ & 1043.086 & 3 & 1080 & 1064.043 & 0.000 & Not significant\\ \hline
$\Lambda_g$ & SNe Ia+$H(z)$+SH0ES $H_0$ & 1044.451 & 4 & 1080 & 1072.393 & 7.333 & Decisive \\ \hline
$\Lambda_1$ ($\lambda_{*} = 0$) & SNe Ia+$H(z)$+SH0ES $H_0$ & 1046.570 & 3 & 1080 & 1067.527 & 2.067 & 
Weak \\ \hline
$\Lambda_2$ ($\beta = 0$) & SNe Ia+$H(z)$+SH0ES $H_0$ & 1044.503 & 3 & 1080 & 1065.460 & 0.000 &  Not significant\\ \hline
$\Lambda_3$ ($\alpha = 0$) & SNe Ia+$H(z)$+SH0ES $H_0$ & 1044.507 & 3 & 1080 & 1065.464 & 0.004 & Not significant\\ \hline
\end{tabular}
\caption{BIC values for the different analysed models, with different $H_0$ priors. $\Delta$BIC was calculated in comparison with model $\Lambda_3$ for each prior, as explained on the text.}
\label{tabBIC}
\end{table}

We may see that the analysis with the Planck prior over $H_0$ indicates strong evidence against model $\Lambda_1$, while the SH0ES prior indicates moderate evidence against model $\Lambda_1$. Both priors, however, can not distinguish between models $\Lambda_2$ and $\Lambda_3$. We can see that the models $\Lambda_2$ and $\Lambda_3$ have a better fit in the case of the Planck $H_0$ prior than in the case of the SH0ES $H_0$ prior. The situation is inverted, however, in the case of model $\Lambda_1$ because the analysis with the Planck $H_0$ prior discards this model, while the SH0ES $H_0$ prior has only weak evidence against $\Lambda_1$.

\begin{figure}[H] 
\begin{center}
    \includegraphics[width=\textwidth]{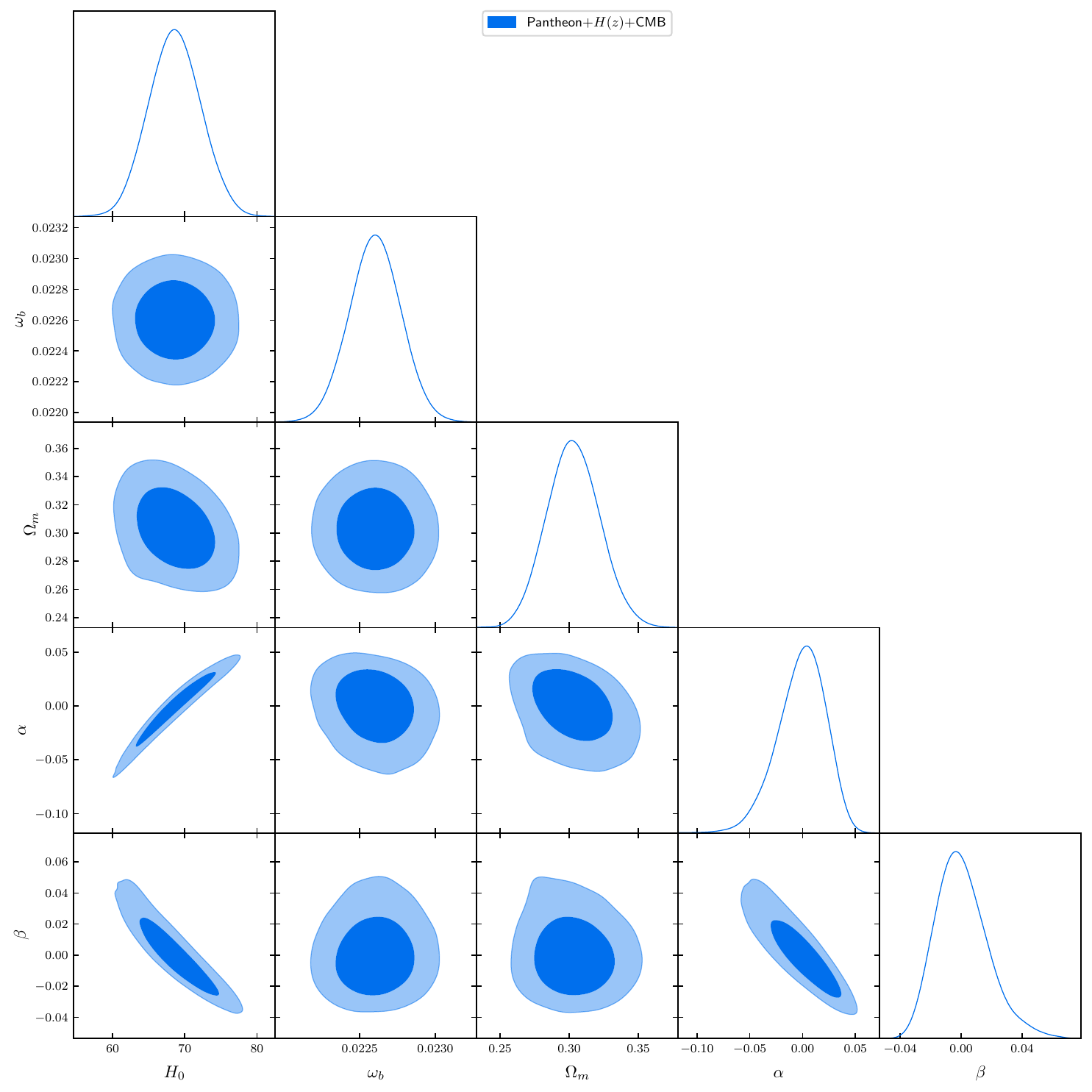}
	\caption{SNe Ia + $H(z)$ + CMB Planck 2018 distance priors constraints for full $\Lambda_g$ model, at 1 and 2$\sigma$ c.l., $H_0$ units are km/s/Mpc.}
 \label{LGeralComb}
\end{center}
\end{figure}

Finally, we have analyzed the full $\Lambda(t)$CDM model, as described by Eq. \eqref{lgeral}. In this case, as it has 1 more free parameter than the subclasses, SNe Ia+$H(z)$+$H_0$ data were not enough to constrain its free parameters. Then, we choose to work with CMB constraints, in combination with SNe Ia Pantheon and $H(z)$ in order to constrain its free parameters. In order to include the CMB constraints we have used the so called ``distance priors'' from Planck, as explained in \cite{ChenEtAl19}. It includes constraints from Planck over quantities like shift parameter $R$, acoustic scale $l_A$ and baryon density $\omega_b\equiv\Omega_bh^2$, where $h\equiv\frac{H_0}{100}$. \cite{ChenEtAl19} present distance priors in the context of 4 models, namely, $\Lambda$CDM, $w$CDM, $\Lambda$CDM+$\Omega_k$ and $\Lambda$CDM+$A_L$. As these priors bring strong constraints from Planck over models which are distinct of $\Lambda(t)$CDM, we choose to work with the prior that yields the weakest constraints, namely, $\Lambda$CDM+$\Omega_k$, weakening the prior bias. In order to speed up the convergence of chains, we have chosen to work with baryon density $\omega_b$ instead of baryon density parameter $\Omega_b$. The results can be seen on Fig. \ref{LGeralComb}.

As can be seen on Fig. \ref{LGeralComb}, there are strong correlations between the parameters $\alpha-H_0$ and $\alpha-\beta$. One can also see that $\alpha$ and $\beta$ are strongly constrained by this analysis. More details can be seen on Tab. \ref{TabCMB}.

\begin{table}[H]
    \centering
        \begin{tabular}{| c | c |c | c | c | c |}
            \hline
            \textbf{Parameter} & \boldmath{$\Lambda_g$} & \boldmath{$\Lambda_2$} ($\beta = 0$) & \boldmath{$\Lambda_3$} ($\alpha = 0$) \\
            \hline 
            \boldmath$H_0$ \textbf{(km/s/Mpc)} & $68.6^{+7.2}_{-7.0}        $ & $68.5^{+3.0}_{-2.7}        $ & $68.6^{+1.7}_{-1.5}        $ \\
            \hline
            \boldmath$\omega_b$ & $0.02260\pm0.00034$ & $0.02260\pm0.00034$ & $0.02261^{+0.00033}_{-0.00034}$ \\
            \hline
            \boldmath$\Omega_m$ & $0.303^{+0.039}_{-0.037}   $ & $0.303^{+0.040}_{-0.038}   $ & $0.303^{+0.036}_{-0.035}   $ \\
            \hline
            \boldmath$\alpha$ & $-0.001^{+0.043}_{-0.047}  $ & $-0.001^{+0.021}_{-0.022}  $ & 0 \\
            \hline
            \boldmath$\beta$ & $0.001^{+0.036}_{-0.033}   $ & 0 & $0.000^{+0.017}_{-0.016}   $ \\
            \hline
        \end{tabular}
    \caption{Constraints over $\Lambda_{g}$, $\Lambda_{2}$ and $\Lambda_{3}$ free parameters with 95\% limits, from SNe Ia+$H(z)$+CMB.}
    \label{TabCMB}
\end{table}

As can be seen on Tab. \ref{TabCMB}, the parameters $\alpha$ and $\beta$, which dictate the $\Lambda$ time dependence, are strongly constrained by CMB, leaving only small windows for $\Lambda$ variation. In fact, we may see that $-0.048<\alpha<0.042$ and $-0.032<\beta<0.037$ at 95\% c.l. This result has to be read with care, as we did not make a full CMB power spectrum analysis in the context of $\Lambda(t)$CDM. We have used, instead, distance priors which depend on models where the DM and DE are separately conserved. With that said, the results for the other parameters are similar to the ones obtained from Planck \cite{Planck2020} in the context of $\Lambda$CDM, where it was found that $H_0=67.4\pm0.5$ km/s/Mpc, $\omega_b=0.0024\pm0.0001$ and $\Omega_m=0.315\pm0.007$. This is expected, as the values obtained for $\alpha$ and $\beta$ are compatible with the $\Lambda$CDM model.

In Tab. \ref{TabCMB} we show, for completeness, the results for the subclasses of models which include a bare cosmological constant $\lambda_*$ term, $\Lambda_2$ and $\Lambda_3$. As it can be seen, the parameters are in general more constrained in the subclasses than in the general model, which is expected, as they have less free parameters. Again, these results shall be taken with care, as we have used an approximated treatment of the CMB results, but we may conclude that the interaction terms are quite constrained in the context of this analysis.

\section{\label{sec: conclusion}Conclusion}
We have analyzed 3 classes of $\Lambda(t)$CDM models against observations of SNe Ia, cosmic chronometers, and priors over the Hubble constant from Planck and SH0ES. We may conclude that 1 class of models, namely, $\Lambda_1$, may be discarded by this analysis, mainly in the case of the Planck $H_0$ prior. In the case of the SH0ES $H_0$ prior, there is weak evidence against this. Models $\Lambda_2$ and $\Lambda_3$ can not be distinguished by this analysis.

At this point, it is important to mention that $\Lambda_1$ is the only considered model that does not have the standard $\Lambda$CDM model as a particular case, once that $\lambda_*=0$. It seems that the data disfavor $\Lambda(t)$CDM models with this feature, at least for the classes of models analyzed here.

However, as one may see in Tables \ref{tabPlL1}-\ref{tabSHL3} and Figs. \ref{InteractionPL} and \ref{InteractionSH}, the current analysis does not discard the possibility of an interaction between pressureless matter and vacuum. We also have seen in the current analysis, that the decaying of pressureless matter into a vacuum is in general favoured against vacuum decay, at least for the recent evolution of the Universe. The $\Lambda_1$ model is the only one where this situation changes in the past, allowing for a decay of the vacuum into pressureless matter.

This would indicate an obstacle for the classes of $\Lambda(t)$CDM models analysed here in order to alleviate the cosmological constant problem. On the other hand, a decaying of matter into vacuum may explain why only recently the vacuum density has become non-negligible.

As a final analysis, in the case of the more general model, $\Lambda_g$, it was necessary to combine SNe Ia+$H(z)$ with CMB in order to constrain its free parameters. In this case, the $\Lambda$ time-dependence was quite constrained, but, as we have used an approximate method, a full analysis with the CMB power spectrum is needed in order to give the final verdict about this model.

We emphasize that in the present article, we have assumed the EoS of vacuum to be exactly $w_{vac} = -1$. However, a recent result for the RVM is that the EoS of vacuum evolves with the cosmic history \cite{Pulido2022b}. This would change our results and may be considered in future works.

Further analysis, considering other observational data, such as BAO, growth factor and full CMB power spectrum, in the lines of \cite{Peracaula2023}, for instance, in order to better constrain these models should be done in a forthcoming issue.

\begin{acknowledgments}
SHP acknowledges financial support from  {Conselho Nacional de Desenvolvimento Cient\'ifico e Tecnol\'ogico} (CNPq)  (No. 303583/2018-5 and 400924/2016-1). This study was financed in part by the Coordena\c{c}\~ao de Aperfei\c{c}oamento de Pessoal de N\'ivel Superior - Brasil (CAPES) - Finance Code 001.
\end{acknowledgments}



\end{document}